\documentclass[twocolumn, showpacs, english, preprintnumbers, amsmath, amssymb, superscriptaddress, aps, prl,longbibliography,nofootinbib]{revtex4-2}
\usepackage{silence}
\usepackage{xspace}

\usepackage[utf8]{inputenc}
\usepackage{graphicx}
\usepackage{grffile}
\usepackage{amssymb}
\usepackage[usenames,dvipsnames]{xcolor}
\usepackage{amsmath,bm}
\usepackage[normalem]{ulem}
\usepackage{appendix} 
\definecolor{orange}{rgb}{1,0.5,0}
\definecolor{goodgreen}{rgb}{0.1,0.5,0}
\definecolor{goodred}{rgb}{0.7,0,0}

\renewcommand\vec{\boldsymbol}
\usepackage{comment}

\usepackage{tikz}
\makeatletter

\usepackage[colorlinks,urlcolor=goodgreen,citecolor=blue,linkcolor=goodred]{hyperref}

\newcommand{\orcid}[1]{\href{https://orcid.org/#1}{\includegraphics[width=8pt]{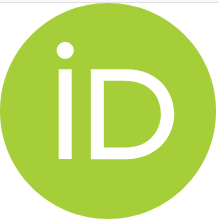}}}

\usepackage{color}

\definecolor{TTH-color}{rgb}{0.0,0.0,1}

\definecolor{SB-color}{rgb}{0.0,0.5,0}

\usepackage{float}

\let\oldepsilon\epsilon \let\epsilon\varepsilon \let\varepsilon\oldepsilon

\newcommand\commastsym{\raisebox{-.5ex}{\shortstack{%
  \(\circ\)\\[-.5ex]%
  \(,\)}}%
}

\makeatother

\usepackage{babel}

\makeatother

\begin{document}
\title{Nonequilibrium spin-splitter effect in altermagnet superconductor hybrids}
\author{Tim Kokkeler \orcid{0000-0001-8681-3376}}
\email{tim.h.kokkeler@jyu.fi}
\affiliation{Department of Physics and Nanoscience Center, University of Jyväskylä, P.O. Box 35 (YFL), FI-40014 University of Jyväskylä, Finland}

\author{Tero T. Heikkilä \orcid{0000-0002-7732-691X}}
\email{tero.t.heikkilä@jyu.fi}
\affiliation{Department of Physics and Nanoscience Center, University of Jyväskylä, P.O. Box 35 (YFL), FI-40014 University of Jyväskylä, Finland}

\author{F. Sebastian Bergeret\orcid{0000-0001-6007-4878}}
\email{fs.bergeret@csic.es}
\affiliation{Centro de Física de Materiales (CFM-MPC) Centro Mixto CSIC-UPV/EHU,
E-20018 Donostia-San Sebastián, Spain}
\affiliation{Donostia International Physics Center (DIPC), 20018 Donostia--San
Sebastián, Spain}
\begin{abstract}
    We study the nonequilibrium spin-splitter effect in superconducting altermagnets and superconductor altermagnet hybrids by computing the alternating spin current and edge the spin density  in the presence of an alternating electric field. We show that while in the normal state the effect is not sensitive to the field frequency, in the superconducting state, there is a strong effect for frequencies on the scale of $\Delta_0$ or lower. We contrast the effect to the spin accumulation induced by the spin-Hall effect, by showing that for the altermagnet spin-splitter effect the out-of-phase spin density does not diverge in the adiabatic limit. This difference is attributed to the absence of any equilibrium spin-splitter effect in altermagnets. In fact, the out-of-phase component vanishes below the gap excitation frequency $2\Delta_0$, because below this frequency the absence of dissipation and the behavior of the system under time-reversal directly determine the relative phase between the charge current, spin current, and spin accumulation. The nonequilibrium effect can be tuned by external parameters like temperature. In fact, it has a nonmonotonic temperature dependence, taking its largest value for temperatures around $0.8T_{c}$. The value at this temperature can be significantly larger than the normal state spin density or the low temperature spin density. Thus, besides using the nonequilibrium spin-splitter effect to identify altermagnets, its tunability makes it also suitable for applications.
\end{abstract}
\maketitle

\section{Introduction}
Altermagnets have been a focal point within condensed matter physics since their recent introduction  \cite{sinova2022emerging,smejkal2022beyond,jungwirth2024altermagnets,reimers2024direct,fedchenko2024observation,bai2024altermagnetism,das2023transport,zhang2024electric,gomonay2024structure,gonzalez2021efficient,fukaya2025josephson,lu2024josephson,zhao2025orientation,krempasky2024altermagnetic,zeng2024observation,ding2024large}. They combine the spin-splitting of the electronic band structure that is typical of ferromagnets, with the absence of a net exchange field, like antiferromagnets. This combination makes their interplay with superconductivity intriguing, since they allow for the investigation of spin-splitting responses \cite{gonzalez2021efficient} without the detrimental responses to superconductivity that are inherent to a net exchange field. This makes them very suitable for extensions within the field of superconducting spintronics, which traditionally combines superconductivity and ferromagnetism \cite{buzdin2005proximity,bergeret2005odd,eschrig2011spin,linder2015superconducting}.

The nonuniform spin-splitting of altermagnets in momentum space allows for the existence of both longitudinal and transverse spin-related responses when a current flows through a superconductor, a feature that has been studied a lot in the context of spin-orbit coupling \cite{edelstein1995magnetoelectric,fujimoto2005magnetoelectric,konschelle2015theory,bergeret2016manifestation,huang2018extrinsic, amundsen2024colloquium,bobkova2017quasiclassical}.  While charge–spin conversion in altermagnets and materials with strong spin–orbit coupling bears significant similarities, there are also important qualitative differences between them. First of all, altermagnetism is nonrelativistic~\cite{sinova2022emerging}, and hence these types of effects are expected to be larger. Secondly, altermagnetism breaks time-reversal symmetry, while spin-orbit coupling does not. This has important consequences on the symmetries of several quantities, such as the conductivity tensor \cite{gonzalez2021efficient}. This difference becomes even more apparent in the presence of superconductivity. Indeed, spin-orbit coupling can create an interfacial spin from a constant charge current both in the normal and superconducting state. On the other hand, for superconductor altermagnet hybrid structures, it has been shown that the symmetries of the observables and parameters under time-reversal imply that a constant supercurrent cannot create a spin accumulation to the interface in altermagnets \cite{kokkeler2025quantum}. This means that to create a spin accumulation in a superconducting altermagnet, either a time dependent input or dissipation is needed. In superconductors, both can be achieved by applying an AC electric field.
\begin{figure}
    \centering
    \includegraphics[width=8.6cm]{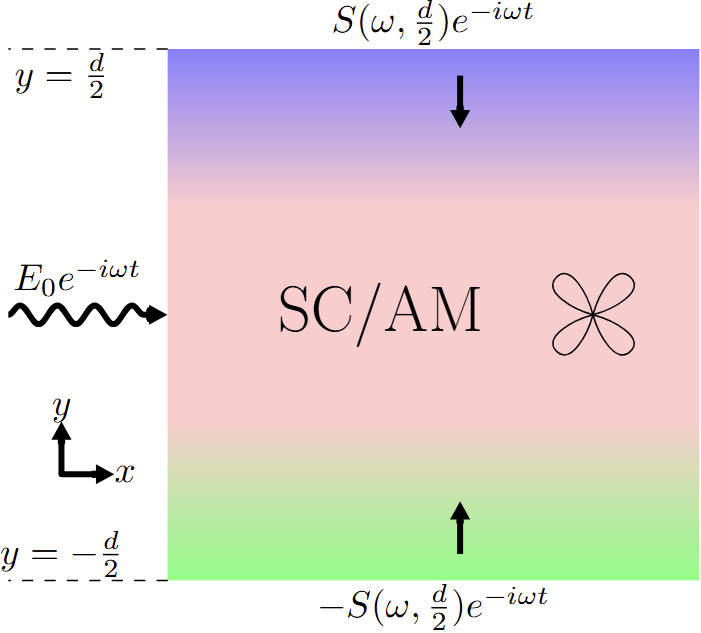}
    \caption{The studied geometry. In the presence of an alternating electric field, an alternating spin polarization is generated at the transverse edges of an altermagnet unless the electric field direction is aligned with the lobes of the spin splitting.
    }
    \label{fig:Setup}
\end{figure}

In this manuscript, we study the generation of a spin accumulation in a superconductor by an
alternating electric field. To this end, we consider a single superconducting altermagnet in the presence of an AC electric field, as illustrated in Fig.~\ref{fig:Setup}. We choose a gauge in which the scalar potential vanishes, so that the field is described via a time-dependent vector potential. We assume that the Fermi velocity is much smaller than the speed of light in the material so that we may ignore relativistic effects, and the spatial dependence of the vector potential on the scale of the coherence length. We show that a spin accumulation appears and it has both a component that is in-phase with the applied field and a component that is out-of-phase. We contrast our results with the spin induced by the spin-Hall effect by showing that below the excitation gap the spin induced by the spin-splitter effect is in-phase with the electric field, while the spin induced by the spin-Hall effect is proportional to the current and hence out of phase with the electric field. We explain that this difference is a consequence of the time-reversal odd nature of altermagnetism. Moreover, we study the parameter dependence of the nonequilibrium spin-splitter effect and show that it is a useful tool to indicate the simultaneous presence of superconductivity and altermagnetism and can thus be used to prove the presence of the altermagnetic state. 

The manuscript is structured as follows. First, we lay out the formalism for the calculation of AC responses in altermagnets and use this to calculate the currents and spin-currents in a bulk superconducting altermagnet in the presence of AC fields. Next, we consider a half-plane geometry and consider the spin accumulation at the edge when applying an AC field for different temperatures. Then, we compare these results with those obtained using a material that exhibits the spin Hall effect, and elaborate on how to distinguish between them.

Throughout, we use a complex electric field $E_0 e^{-i\omega t}$, where $E_0$ is taken to be real. For this reason the currents and spin accumulations are also complex. To obtain physical observables, the currents and spins computed at $\pm \omega$ need to be summed. Thus, physical observables are real if and only if the currents and spin accumulations at $\pm\omega$ are related by complex conjugation. In this case the real part corresponds to the component in-phase with the electric field ($\cos{\omega t}$) and the imaginary part with the out-of-phase component ($\sin{\omega t}$). We show in Appendix \ref{sec:RealObservables} that our equations indeed obey this feature. Thus, all observables can be determined from the solutions for positive frequencies. Therefore, in the rest of the manuscript we focus on  $\omega>0$.  We use units with $e = c = \hbar = 1$.

\section{Transport equation for Altermagnets}
In this work,  we focus on the spin currents and spin accumulations within altermagnets in the presence of superconductivity. These can be described using the kinetic equation of the material, the  Usadel equation \cite{usadel1970generalized}, which describes the evolution of the dirty limit quasiclassical Green's function, that is, the momentum-averaged Green's function. Because we study nonequilibrium effects, we employ the Keldysh formalism, which describes the retarded and advanced Green's functions, as well as the distribution function. They are combined into a matrix space, the Keldysh-Nambu-spin space, in which the Green's function takes the following form:
\begin{align}
    g(t_1,t_2) &= \begin{bmatrix}
        g^R(t_1,t_2)&g^K(t_1,t_2)\\0&g^A(t_1,t_2)
    \end{bmatrix}\;.
\end{align}
Here $g^{R,A}$ are the retarded and advanced quasiclassical Green's functions in Nambu-spin space, while the Keldysh Green's function can be written as $g^K = g^R\circ h-h\circ g^A$, where $h$ is a matrix distribution function in Nambu-spin space, and $\circ$ denotes convolution, that is, $X\circ Y (t_1,t_2) = \int_{-\infty}^{\infty} dt_{3} X(t_1,t_3)Y(t_3,t_2)$.

The Usadel equation takes a general form in terms of the matrix current $\vec{\mathcal{J}}$ of the system, a position, time and energy dependent matrix in Keldysh-Nambu-spin space that contains for example the charge current and spin currents as elements, and the corresponding matrix torque $\mathcal{T}$ for these currents: 
\begin{align}
    \vec{\hat{\partial}}\cdot\vec{\mathcal{J}} = [\hat{\omega}_{t_1,t_2}\tau_3+\Delta\tau_2\commastsym g]+\mathcal{T}\;.\label{eq:UsadelAMCT}
\end{align}
Here $g$ is the quasiclassical Green's function in Keldysh-Nambu-spin space, $\vec{\hat{\partial}}\cdot = \vec{\partial}\cdot -i[\vec{A}(t_1)\delta(t_1-t_2)\tau_3\commastsym\cdot]$ is the covariant derivative, $\vec{A}$ is the vector potential, $\hat{\omega}_{t_1,t_2} = \delta(t_1-t_2)\partial_{t_{1}}$, and $\Delta$ is the pair potential, which obeys the BCS self-consistency relation. The Pauli matrices in Nambu space are indicated by $\tau_i, i = 1,2,3$, and $[\cdot\commastsym\cdot]$ denotes the commutator with respect to the convolution $\circ$. 
Here we use a specific type of the Usadel equation to describe altermagnets. Since we are interested in nonrelativistic effects, we ignore spin-orbit coupling. In that case, the matrix current and torque in an altermagnet read \cite{kokkeler2025quantum}:
\begin{align}
    \mathcal{J}_k &=-Dg\circ\hat{\partial}_{k}g+\frac{D}{4}T_{jk}\{\tau_{3}\sigma_{z}+g\circ\tau_{3}\sigma_{z}g\commastsym  g\circ\hat{\partial}_{j}g\}\nonumber\\&+i\frac{D}{4}K_{jk}[\tau_{3}\sigma_{z}+g\circ\tau_{3}\sigma_{z}g\commastsym \hat{\partial}_{j}g]\;,\label{eq:CurrentAM}\\
    \mathcal{T} &=T_{jk}[\tau_3\sigma_z,\hat{\partial}_j g\circ\hat{\partial}_k g]+iK_{jk}[\tau_3\sigma_z,g\circ\hat{\partial}_j g\circ\hat{\partial}_k g]\nonumber\\&+\Gamma_{ab}[\tau_3\sigma_a g\tau_3\sigma_b\commastsym g]\;.\label{eq:TorquesUsadelAM}
\end{align}
 Here $\{\cdot\commastsym\cdot\}$ denotes an anticommutator with respect to the convolution $\circ$, $D$ is the spin-averaged diffusion constant of the system, $\sigma_a, a = 1,2,3$ denote the Pauli matrices in spin space, and $K_{jk},T_{jk}$ are symmetric tensors whose magnitude depend on the microscopic parameters of the system such as the spin-splitting of the bands and the scattering rate. Their form is determined by the crystal symmetries of the material, which can be described using the spin space groups \cite{smejkal2022beyond}. They are only nonzero for altermagnets of the $d$-wave type. 
Which elements are nonzero depends on the choice of axis. In 2D, if we choose the axes corresponding to the directions in which the spin-splitting is maximal, $T_{xx} = -T_{yy}$ and $K_{xx} = -K_{yy}$ are the only allowed elements. However, if they correspond to a direction without spin splitting, $T_{xy} = T_{yx}$ and $K_{xy} = K_{yx}$ are the only nonzero elements. Because in this paper we are interested in transverse effects, we opt for the latter case and consider electric fields in the $x$-direction. Lastly, $\Gamma_{ab}$ is the spin-relaxation tensor, which appears in all types of altermagnetism and has several  contributions, for example due to a mechanism similar to the Dyakonov-Perel mechanism \cite{vasiakin2025disorder}, due to magnetic impurities \cite{lamacraft2000tail,lamacraft2001superconductors,marchetti2002tail}, and due to the mixing of spin by scattering that has been shown to appear in antiferromagnets \cite{fyhn2023quasiclassical}. The presence of $\tau_3$'s in this term reflect the fact that the spin-relaxation is of magnetic nature, and not caused by spin-orbit coupling. In collinear altermagnets it has two independent components, the spin relaxation rate for spins along the collinear axis and the spin relaxation rate for spins perpendicular to the collinear axis. 

The observables can be extracted from the Green's function $g$ and the matrix current $\mathcal{J}$. Indeed, we define the charge current $j_k$ and the spin current $j_k^a$ at a time-instant $t$ as components of the matrix current in Eq.~(\ref{eq:CurrentAM}), evaluated at equal times:
 \begin{align}
     j_k(t)&=\frac{\pi\nu_0}{4}\text{tr}\Big(\rho_1\tau_3J_k (t,t)\Big)\;, \label{eq:ChargeCurrentDefinition}\\
     j_k^a(t) &= \frac{\pi\nu_0}{4} \text{tr}\Big(\rho_1\sigma_a J_k(t,t)\Big)\;,\label{eq:SpinCurrentDefinition}
 \end{align}
where $\nu_0$ is the density of states per spin of the material, and $\rho_1$ is the first Pauli matrix in Keldysh space. 

The corresponding time-dependent change in the spin density (in direction $a$) are
\begin{align}
    S^a (t) &= \frac{\pi\nu_0}{4}\text{tr}\Big(\rho_1\tau_3\sigma_ag(t,t) \Big)\label{eq:Spinaccumulation} \;.
\end{align}

We calculate the spin current and spin responses for an alternating electric field $\vec{E}$ both in the regime where the frequency is below and above the excitation gap of the superconductors. Below the excitation gap, there is no dissipation, that is, the system is reversible. In this case the time-reversal operator fixes the relative phase between the different observables. For example, as predicted by the Mattis-Bardeen theory \cite{mattis1958theory}, the time-reversal odd longitudinal charge current is out-of-phase with the time-reversal even electric field for dissipationless transport. For the spin-splitter effect, we may write
\begin{align}
    j_{k}^a (\omega) = \chi^{(1)}_{ajk}(\omega,T_{jk}) E_j(\omega)\;,
\end{align}
where by spin-inversion symmetry, the response function satisfies $\chi^{(1)}_{jk}(\omega,T_{jk}) = -\chi^{(1)}_{jk}(\omega,-T_{jk})$. Since under time-reversal, $\omega$ and $T_{jk}$ change sign, while $\vec{E}$ and $j_k^{a}$ do not, we conclude that in the absence of dissipation, 
\begin{align}
    \chi^{(1)}_{ajk}(\omega,T_{jk}) = \chi^{(1)}_{ajk}(-\omega,-T_{jk}) = -\chi^{(1)}_{ajk}(-\omega,T_{jk})\;,
\end{align}
that is, the in-phase component of the transverse spin-current requires dissipation. This restriction is opposite to the one found for the spin-Hall effect \cite{hijano2023dynamical}, because the spin-Hall tensor, in contrast to the altermagnet tensor, is time-reversal even. This difference between the two systems can also be understood using Onsager's relations \cite{hijano2023dynamical}, exploiting that the altermagnet tensor is symmetric in its spatial indices, while the spin-Hall tensor is anti-symmetric.

For the generated spin we have
\begin{align}
    S^a (\omega) = \chi^{(2)}_{ajk}(\omega,T_{jk}) E_j(\omega)\;,
\end{align}
where also the response function $\chi^{(2)}$ is odd under spin-inversion symmetry. Since $S^a$ is odd under the time-reversal operator, we find
\begin{align}
    \chi^{(2)}_{ajk}(\omega,T_{jk}) = -\chi^{(2)}_{ajk}(-\omega,-T_{jk}) = \chi^{(2)}_{ajk}(-\omega,T_{jk})\;,
\end{align}
that is, the spin is in-phase with the electric field and out-of-phase with the current in the absence of dissipation. In the limit $\omega\xrightarrow{}0$ this latter restriction implies the absence of the equilibrium spin-splitter effect, as found in \cite{kokkeler2025quantum}.

In the following sections we solve the Usadel equation for altermagnets, Eq.~(\ref{eq:UsadelAMCT}) to first order in the vector potential $\vec{A}$ to compute observables.
\section{Spin-splitter conductivity}

An AC electric field  modifies the vector potential to $A_{x}(t) = \frac{E_{0}}{\omega}e^{-i\omega t}$ and it enters the matrix current via the covariant derivative. In a homogeneous system $\hat{\partial}_{x}\cdot\xrightarrow{}-i[\tau_{3}A_{x}(t_1)\delta(t_1-t_2),\cdot]$. 
In that case the charge current and the spin current are also harmonic, that is, $j_k(t) = j_k(\omega)e^{-i\omega t}$ and $j_k^{a}(t) = j_k^{a}(\omega)e^{-i\omega t}$. From this we compute the longitudinal charge conductivity $\sigma_{xx}$ and the transverse spin conductivity $\sigma_{xy}^{a}$, which we refer to as the spin-splitter conductivity $\sigma_{SSE}$ to distinguish it from the spin-Hall conductivity $\sigma_{SHE}$, as
\begin{align}
    \sigma_{xx}(\omega) &= \frac{j_x(\omega)}{E_{0}}\;,\\
    \sigma_{SSE}(\omega) &= \frac{j_{y}^{a}(\omega)}{E_0}\;.
\end{align}

In this section,  we consider a planar superconductor without boundaries. 
We keep only terms of the first order in $E_{0}$ in the current. Since $\vec{A}$ is in the $x$-direction, there is a longitudinal current from the usual diffusion term, i.e., the first term in Eq.~(\ref{eq:CurrentAM}), and possibly a transverse spin current in the $y$-direction due to the off-diagonal $T_{jk}$ and $K_{jk}$ terms. We focus on these latter terms in Eqs.~(\ref{eq:UsadelAMCT}-\ref{eq:TorquesUsadelAM}). 

Importantly, because altermagnets are inversion symmetric, the Usadel equation does not contain terms with one derivative. Therefore, there the bulk equation does not contain any term to first order in $\vec{A}$ without derivatives. This means that to compute the first order response, we only require the zeroth order Green's function. It is of   the bulk BCS form, and hence it only depends on the time difference $t_1-t_2$. The Fourier transform of its retarded part reads
\begin{align}
    g^R(E) = \frac{1}{\sqrt{\Delta^2-(E+i\eta)^2}}(-i(E+i\eta)\tau_3+\Delta\tau_2) \label{eq:BCSg}\;.
\end{align}
Here, $\Delta$ is the superconducting gap and $\eta$ is the Dynes parameter \cite{dynes1978direct}, which is often introduced to regularize the Green's function at $E = \Delta$, but it also roughly accounts for inelastic effects \cite{mikhailovsky1991thermal}.  The advanced part is related to the retarded part via $g^A(E) = -\tau_3 (g^R(E))^{\dagger}\tau_3$ and the Keldysh part is $g^K(E)=(g^R(E)-g^A(E))h_0$, where $h_0(E) = \tanh{\frac{E}{2k_BT}}$ reflects the Fermi-Dirac distribution of electrons in equilibrium.

Currents are first order in $\vec{A}$ and can be expressed in terms  of the equilibrium Green's function $g$. The latter  has no spin dependence. 
Therefore it commutes with $\sigma_z$, which means we may simplify the expression for the spin current,  that follows from Eq.~(\ref{eq:CurrentAM}) to $j_{y}^{x} = j_{y}^{y} = 0$ and
\begin{widetext}
\begin{align}
    j_{y}^{z}(t) & = \frac{\sigma_D}{16\omega}PT_{xy}\Bigg(\text{tr}\Big(\rho_{1}\Big\{(\tau_{3}+g\circ\tau_{3}g)\commastsym g\circ[E_{0}e^{-i\omega t_1}\delta(t_1-t_2)\tau_{3}\commastsym g]\Big\}\Big)\nonumber\\&+i\frac{\sigma_D}{16\omega}PK_{xy}\text{tr}\Big(\rho_{1}\Big[(\tau_{3}+g\circ\tau_3g)\commastsym [E_{0}e^{-i\omega t_1}\delta(t_1-t_2)\tau_{3}\commastsym g]\Big]\Big)\Bigg)_{t_1 = t,t_2 = t}\;,
\end{align}
\end{widetext}
where $\sigma_D = 2\nu_0 D$ is the Drude conductivity of the material.
We may simplify this expression by noting that the last term vanishes identically. Indeed, it has been shown before that this term merely leads to spin-precession, and therefore cannot create a spin current from a charge current \cite{kokkeler2025quantum}. Next, to evaluate this expression, we use a strategy similar to the one introduced in \cite{hijano2023dynamical} for the spin-Hall effect. The details of the derivation can be found in the Supplemental Material; here, we present the results.  After a Wigner transform and keeping leading order terms in the gradient expansion \cite{larkin1975nonlinear,kopnin2001theory}, we obtain a compact expression for the spin current:
\begin{widetext}
\begin{align}
    j_{y}^{z}(t) = j_{y}^{z}(\omega)e^{i\omega t} &= \frac{\sigma_D}{16\omega}PT_{xy}E_{0}e^{-i\omega t}\int_{-\infty}^{\infty} dE \text{tr}\Bigg(\rho_{1} \Big(\tau_{3}+g(E)\tau_{3}g(E)\Big)g(E)\Big(\tau_{3}g(E+\omega)-g(E)\tau_{3}\Big)\nonumber\\&+\rho_1 g(E)\Big(\tau_{3}g(E+\omega)-g(E)\tau_{3}\Big)\Big(\tau_{3}+g(E+\omega)\tau_{3}g(E+\omega)\Big)\Bigg)\;.\label{eq:spincurrentExpression}
\end{align}
\end{widetext}
\begin{figure*}
    \centering
    \includegraphics[width=8.6cm]{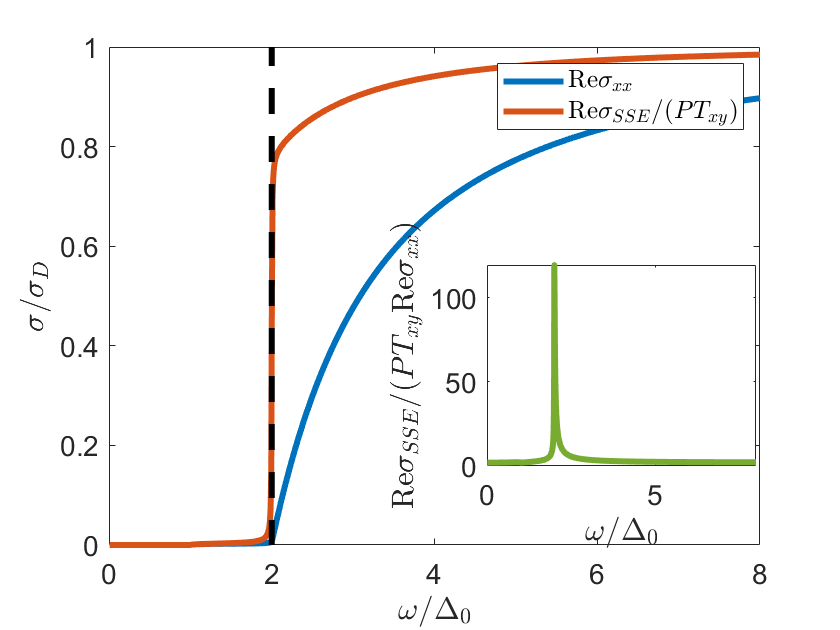}
    \includegraphics[width=8.6cm]{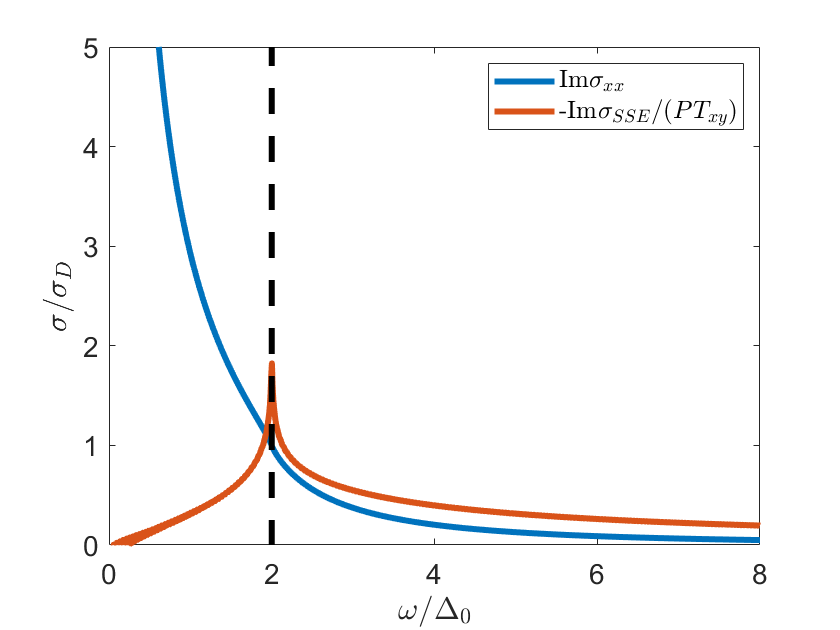}
    \includegraphics[width=8.6cm]{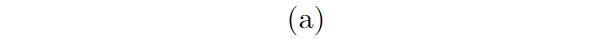}
    \includegraphics[width=8.6cm]{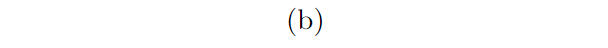}
    \caption{The real (a) and the imaginary (b) parts of the zero temperature spin-splitter conductivity and the longitudinal conductivity.  The longitudinal conductivity follows the usual Mattis-Bardeen theory. The spin-splitter conductivity behaves similarly to the longitudinal charge conductivity, but it goes up much faster near $\omega = 2\Delta_0$, while there longitudinal component vanishes rather than diverges as $\omega\xrightarrow{}0$. The dashed line indicates the excitation gap, below which dissipation only appears due to the Dynes parameter  $\eta = 0.002\Delta_0$. \textit{Inset:} Normalized ratio of the spin-splitter conductivity and longitudinal charge conductivity $ \sigma_{SSE}/(\sigma_{xx}P T_{xy})$.}
    \label{fig:realandimag}
\end{figure*}

We first analyze  this expression in the normal state, that is, $\Delta= 0$. In that case, $g^R(E) = -g^A(E) = \tau_3$, independent of energy, while $g^K(E) = 2\tau_3 h_0(E)$. Thus, Eq.~(\ref{eq:spincurrentExpression}) implies that in the normal state the charge current and the transverse spin current, and hence the longitudinal charge conductivity and spin-splitter conductivity, are related via
\begin{align}
    j_{y}^{z}(\omega)  &=PT_{xy}j_{x}(\omega)\;,\\
    \sigma_{SSE}(\omega) &= PT_{xy}\sigma_{xx}(\omega)\;.
\end{align}
That is, the ratios of the transverse spin and longitudinal charge responses are independent of frequency and equal $PT_{xy}$. This is the spin-splitter effect predicted in Ref. \cite{gonzalez2021efficient}, generalized to disorder systems. On the other hand, the charge current is given by 
\begin{align}
    j_{x}(\omega) &= \frac{\sigma_DE_0}{16\omega}\int_{-\infty}^{\infty}\hspace{-0.35cm}dE \text{tr}\Bigg(\rho_{1}\tau_{3}g(E)\Big(\tau_{3}g(E+\omega)-g(E)\tau_{3}\Big)\Bigg) \nonumber\\&= 
    \frac{\sigma_D}{2\omega}E_0\int_{-\infty}^{\infty}dE \Big(h_{0}(E+\omega)-h_{0}(E)\Big) = \sigma_D E_0\;,
\end{align}
which recovers Ohm's law.

In the superconducting case, $\Delta = \Delta_0>0$, we evaluate Eq. (\ref{eq:spincurrentExpression}),  numerically. The results  are shown in  Fig.~\ref{fig:realandimag} for the real (a) and imaginary (b) parts of the longitudinal charge conductivity and spin-splitter conductivity at zero temperature.
The longitudinal response follows the Mattis-Bardeen theory \cite{mattis1958theory}, with an out-of-phase component that diverges at $\omega = 0$, corresponding to the existence of a supercurrent, and decreases as $\omega$ increases. Below the excitation gap $\omega = 2\Delta_0$, and in the absence of inelastic processes,  there is no dissipation, and hence the charge current must be out-of-phase with the vector potential. 
In the case of a finite Dynes parameter [Eq.~(\ref{eq:BCSg})], the in-phase component does not exactly vanish for $\omega < 2\Delta_0 = 2\Delta(T = 0)$, but it remains small compared to the normal-state response due to the in-gap density of states, which leads to low-frequency dissipation.
 
Next we consider the spin-splitter conductivity. Like the longitudinal charge conductivity, far above $\omega = 2\Delta_0$ the real part converges to the normal-state value, which as discussed above equals $PT_{xy}\sigma_D$. 
Meanwhile below $2\Delta_0$ the real part of the spin-splitter conductivity goes to approximately zero, with only a contribution due to the finite Dynes parameter remaining. This reflects that for the spin-splitter effect the real part of the spin-splitter conductivity is the dissipative part and thus requires quasiparticles. The imaginary part, which is nondissipative, has a peak at $\omega = 2\Delta_0$ and vanishes both in the low and high frequency limits.

In contrast to the longitudinal charge conductivity, there is no divergence at small $\omega$ for the spin-splitter conductivity $\sigma^z_{xy}$, because of the absence of an equilibrium transverse spin current.
This is generically true in collinear systems \cite{delasheras2025}, 
 even though a conversion between supercurrents and spin supercurrents is allowed by time-reversal symmetry and magnetizations do arise in the material. The reason for this is that the magnetization is purely a quasiparticle effect, and the condensate consists only of singlets and zero-spin projection triplets. This means the condensate cannot carry any spin. 
This explains the absence of a divergence at zero frequency in the response.
Nevertheless, the spin-splitter conductivity does become nonzero below the excitation gap of $2\Delta_{0}$. The appearance of this dissipationless signal is due to the presence of the quasiparticle states for $E\geq2\Delta_0$, which, give rise to a reactive response off-resonance, as required by causality.

For $\omega$  larger than $2\Delta_0$ the spin-splitter conductivity increases sharply and approaches its normal state value in a shorter window than the charge conductivity does. We consider this effect in more detail by plotting their ratio in the inset of Fig.~\ref{fig:realandimag}, which shows a pronounced peak at $\omega = 2\Delta_0$. Indeed, close to the peak, even in the presence of a Dynes parameter of 0.002, the ratio between the two is orders of magnitude larger than $PT_{xy}$, which shows that even if the altermagnet coefficient is comparatively small, the response for $\omega\approx 2\Delta_0$ can still be comparable to or even larger than the longitudinal response.

\section{Spin accumulation}
\begin{figure}
    \centering
    \includegraphics[width=8.6cm]{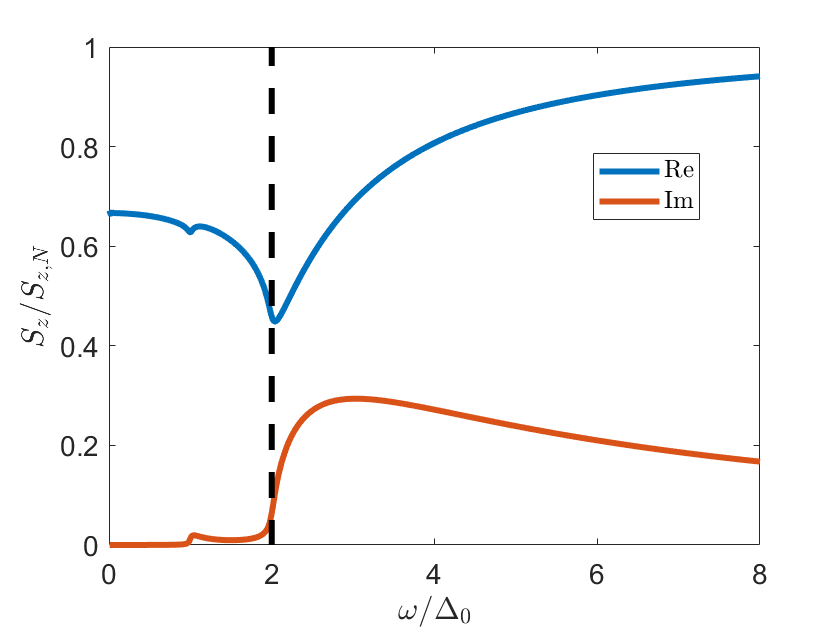}
    \caption{Real and imaginary, i.e., in- and out-of-phase parts of $S(\omega)$ upon application of an electric field $E(\omega)$, normalized to the normal-state spin accumulation $S_{z,N} = \nu_0 PT_{xy}dE(\omega)$ in the limit $d\ll \xi$ and zero temperature. The dashed line indicates the excitation gap, below which any dissipative signal is induced by $\eta = 0.002\Delta_0$.}
    \label{fig:Spin}
\end{figure}

While spin currents are interesting from a fundamental perspective,  they cannot be measured directly. However, in many cases they come along with a spin accumulation at the interface, which is often measured instead. In the normal state the spin current and spin accumulation are closely related and therefore the spin accumulation is an indirect probe of the spin currents in the material. In contrast, in a superconductor this is not the case. Indeed, as was found in the spin-Hall case, even in the absence of a spin current, the spin accumulation at the edges of the sample can be finite in the presence of a charge current \cite{bergeret2016manifestation,huang2018extrinsic}. 
\subsection{Spin accumulation in an infinite strip}
 We consider a setup that is finite in the $y$-direction, see Fig.~\ref{fig:Setup}, and calculate the spin-accumulation via $S^{z}(t,z) = S^{z}(\omega,z)e^{-i\omega t}$ and, following Eq.~\eqref{eq:Spinaccumulation},
\begin{align}
    S^{z}(\omega,z) & =-\frac{\nu_0}{8} \int_{-\infty}^{\infty}dE\text{tr}\Big(\rho_1\tau_3\sigma_zg\Big)\;.
\end{align}
We first consider the case in which the thickness $d$ is much smaller than the superconducting coherence length $\xi = \sqrt{D/\Delta}$.
By setting the current, Eq. (\ref{eq:CurrentAM}) at the boundaries $y = \pm \frac{d}{2}$ to zero we obtain, to lowest order in the thickness, 
\begin{align}
    S^{z}(t,\pm d/2) &\approx \pm\frac{d}{2}\partial_{y}S^{z} = \mp\frac{d}{2}\frac{\nu_0}{8} \int_{-\infty}^{\infty}dE\text{tr}\Big(\rho_{1}\tau_{3}\sigma_{z}\partial_{y}g\Big)\nonumber\\& = \mp\frac{d}{2}\frac{\nu_0}{8} \int_{-\infty}^{\infty}dE\text{tr}\Big(\rho_{1}\tau_{3}\sigma_{z}gJ_{1,x}\Big)\;,\label{eq:Shortlimitspin}
\end{align}
 where we denote $J_{1,x} = DP\Big(T_{xy}\{\tau_{3}\sigma_{z}+g\circ\tau_{3}\sigma_{z}g\commastsym g\circ\hat{\partial}_{x}g\}+iK_{xy}[\tau_{3}\sigma_{z}+g\circ\tau_{3}\sigma_{z}g\commastsym g ]\Big)$. Due to the appearance of an extra factor $\tau_3 g$ compared to Eq.~(\ref{eq:spincurrentExpression}), in the superconducting state the induced spin is not necessarily proportional to, or even in-phase with the spin current, only in the normal state there is a direct connection between the two. This reflects that a time-reversal odd gradient of equilibrium spin does not lead to a time-reversal even spin current in the absence of dissipation.
\begin{figure*}
    \centering
    \includegraphics[width=8.6cm]{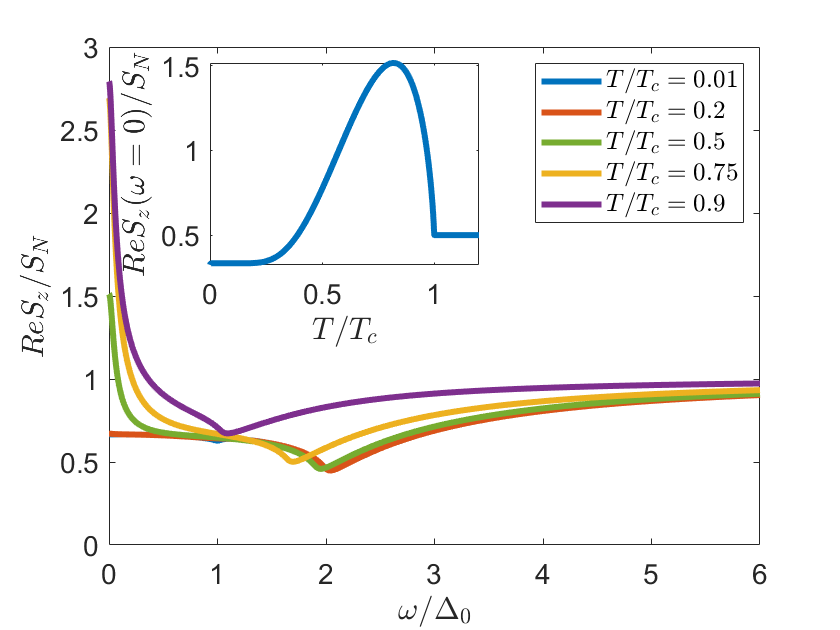}
    \includegraphics[width=8.6cm]{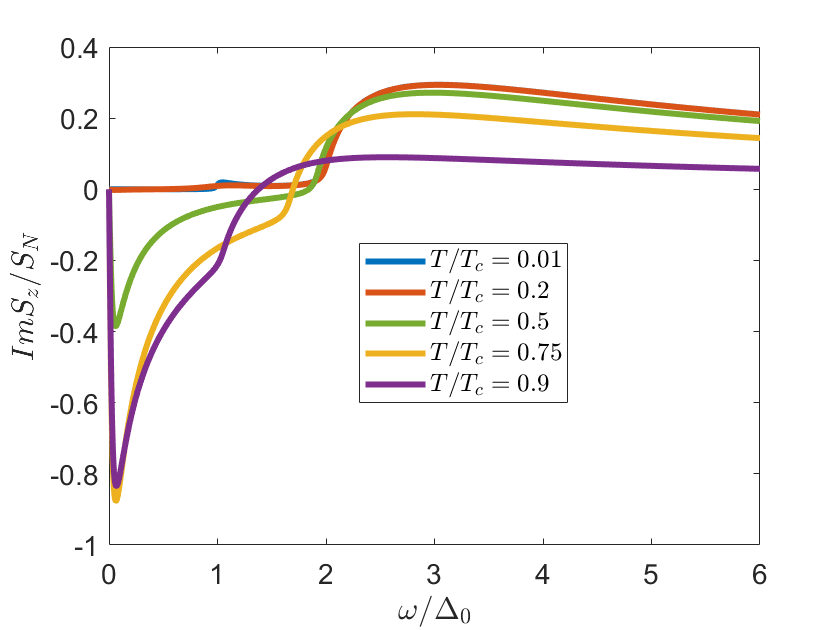}
    \includegraphics[width=8.6cm]{Figures_to_use/A.png}
    \includegraphics[width=8.6cm]{Figures_to_use/B.png}
    \caption{Nonequilibrium spin-splitter effects for different temperatures. As $T$ approaches $T_{c}$, the features induced by superconductivity appear for smaller $\omega$, reflecting the smaller gap, and are more pronounced, reflecting that the Green's function changes faster with energy in the limit $d\ll \xi$. \textit{Inset}: The adiabatic response as a function of temperature. There is a peak close to $T=T_{c}$, in which the adiabatic response exceeds the normal state value significantly. We used $\eta = 0.002\Delta_0$. There are quasiparticles for each frequency because of the nonzero population of states above the gap.}
    \label{fig:Tempdep}
\end{figure*}
As derived in Appendix \ref{sec:SpinAccumulationInterface}, we may write the combination of the two contributions, keeping in mind that $g$ commutes with $\sigma_z$, as:
\begin{align}
    S^{z}(\omega,\pm d/2) & = \mp\frac{\nu_0}{16\omega}E_0 PT_{xy}\frac{d}{2}\int_{-\infty}^{\infty}dE\nonumber\\\text{tr}\Bigg(3\rho_1&  \Big(\tau_3 g(E+\omega)-g(E)\tau_3\Big)\nonumber\\-\rho_1 \tau_3& g(E)\tau_3\Big( g(E+\omega)-g(E)\Big)\tau_3g(E+\omega)\Bigg) \;.\label{eq:Spinacc}
\end{align}
Like for the spin current, the term $K_{jk}$ has no contribution to the spin.
In the normal state, we obtain, as long as $\omega\ll \frac{D}{d^2}$,
\begin{align}
    S^z(\omega,\pm d/2) = \pm S_{z,N} = \mp 2\nu_0dPT_{xy}E_0\;.
\end{align}

Numerical evaluation for $S^z$ in a superconductor gives the results shown in Fig.~\ref{fig:Spin} at zero temperature. For a superconducting altermagnet, there are both an in- and an out-of-phase component of the spin. The in-phase component is finite for any frequency, but it has a minimum at $\omega = 2\Delta_0$ and its maximum at $\omega = 0$ is smaller than the normal state accumulation that is reached for large frequencies. Like the spin-current, the spin below $2\Delta_0$ is not a property of the condensate, but can be understood as an accumulation of quasiparticles that counteracts the bulk quasiparticle spin-current at the boundary. Unlike the spin-current, which vanishes at $\omega\rightarrow 0$,  the spin accumulation is even in $\omega$ and remains finite as $\omega\xrightarrow{}0$. 
Specifically, it converges to $\frac{2}{3}$ of the normal state value, see Fig.~\ref{fig:Spin} and  Appendix \ref{sec:Adiabatic} for details.  
This difference in low-frequency behavior between spin current and spin accumulation at low frequencies appears because the spin current is necessarily odd in frequency, while the spin is necessarily even in frequency.  

The spin-splitter response of altermagnets is in sharp contrast with the AC response due to the spin-Hall effect. For systems with a spin-Hall effect, the generated spin currents have been discussed in \cite{hijano2023dynamical}. It was shown that the spin currents  are in-phase with the applied field. Here, we discuss the generated spins via the spin Hall effect by applying the same methods as for the altermagnet, but now to the equations for a material with finite spin--orbit coupling and, hence, a spin Hall angle $\theta$
\cite{virtanen2021magnetoelectric}. For the spin accumulation at $y=\pm d/2$ we obtain 
\begin{align}
    S^z(\omega, \pm\frac{d}{2}) & = \mp\int_{-\infty}^{\infty}dE\frac{d}{2}\frac{\nu_0}{8}E_0\theta\text{tr}\Bigg(\rho_1\tau_3 g(E)\Big(\tau_3g(E+\omega)\nonumber\\&-g(E)\Big)\Bigg)= \mp\theta\frac{d\nu_0}{2\sigma_D}j_x(\omega)\;.
\end{align}
That is, the spin induced by the spin-Hall angle is directly proportional to the charge current, reflecting the universality of the spin-galvanic effect that was also found for bulk spin galvanic effect \cite{kokkeler2025universal}. In particular, in the presence of a supercurrent there is a finite spin, leading to a divergence of the spin accumulation as $\omega \rightarrow 0$ following the Mattis-Bardeen response.

This is significantly different from the result for spin currents, which are real and do not diverge \cite{hijano2023dynamical}. This difference in behavior of spin currents and spin accumulations bears similarity to the previously discussed altermagnet case, the spin currents and spin have different symmetries under $\omega\xrightarrow{}-\omega$ and therefore necessarily have different behavior in the adiabatic limit. 

\begin{figure*}
    \centering
    \includegraphics[width=8.6cm]{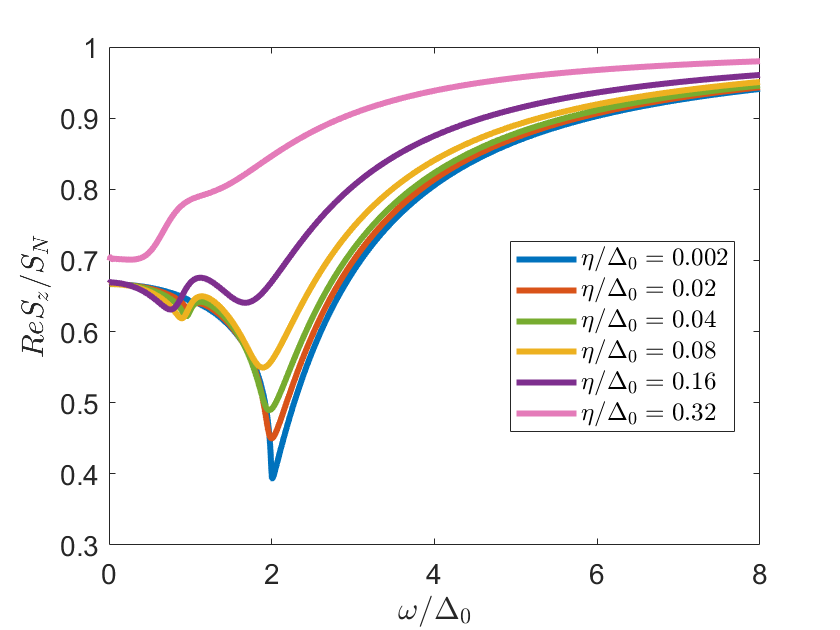}
    \includegraphics[width=8.6cm]{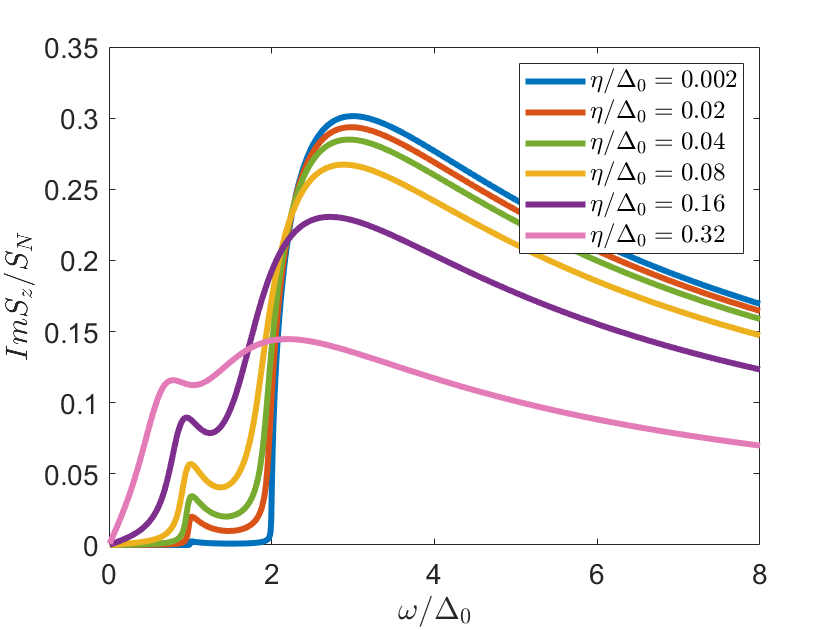}
    \includegraphics[width=8.6cm]{Figures_to_use/A.png}
    \includegraphics[width=8.6cm]{Figures_to_use/B.png}
    \caption{The dependence of the generated spin in the limit $d\ll \xi, l_s$ on the Dynes parameter at $T = 0.1T_c$. The results are normalized using the spin accumulation that appears in the normal state. There is dissipative signal for all frequencies because of the residual density of states induced by the Dynes parameter}
    \label{fig:Etadep}
\end{figure*}
\subsection{Parameter dependence}
Next, we consider how the nonequilibrium spin accumulation depends on several parameters of the system. Specifically, we focus on the temperature, which can be used as an external variable, on the Dynes parameter, which we use to understand the effect of in-gap quasiparticles, and the spin relaxation rate.

In Fig.~\ref{fig:Tempdep} we show the real (a) and imaginary (b) parts of $S(\omega)$ for several different temperatures. To this, end, we choose a nonzero temperature in the distribution function $h_{0}(E) = \tanh{\frac{E}{2k_{B}T}}$ and calculated the gap self-consistently as a function of temperature. For low temperatures, $T\ll T_c$, the results remain similar to the zero temperature result. However, as the temperature approaches the critical temperature, there is a peak at  low $\omega$. For the real part the maximum is at $\omega =0$, while the imaginary part  still vanishes at $\omega = 0$, but has a peak at a finite value of $\omega$. To understand the behavior of the adiabatic response as a function of temperature better, we show in the inset of Fig.~\ref{fig:Tempdep}(a) 
the real part of $S(\omega= 0)$ as a function of temperature. Below the critical temperature the adiabatic response has a gradual dependence on temperature, with a peak around $T\approx 0.82 T_{c}$, and approaching the normal state value as $T\xrightarrow{}T_{c}$. Above $T_{c}$ we recover the normal state behavior, that is, the response is frequency independent.

The results also depend sensitively on the Dynes parameter $\eta$, since next to regularizing the gap of the superconductor it also leads to the presence of in-gap states. This has a large influence on the generated spin, as shown in Fig.~\ref{fig:Etadep}. Via an increase of $\eta$, the generation of an out-of-phase spin, Fig.~\ref{fig:Etadep} (b) below the gap becomes allowed. Nevertheless, it still vanishes in the adiabatic limit, $\omega\xrightarrow{}0$, as it should. The peak also shifts from $\omega \approx \Delta_0$ to $\omega \approx \Delta (\eta,T)$. The influence of $\eta$ on the in-phase component, Fig.~\ref{fig:Etadep} (a) is only qualitative, with all features becoming less sharp and shifting to lower frequencies due to the decrease of the gap.

\begin{figure*}
    \centering
    \includegraphics[width=8.6cm]{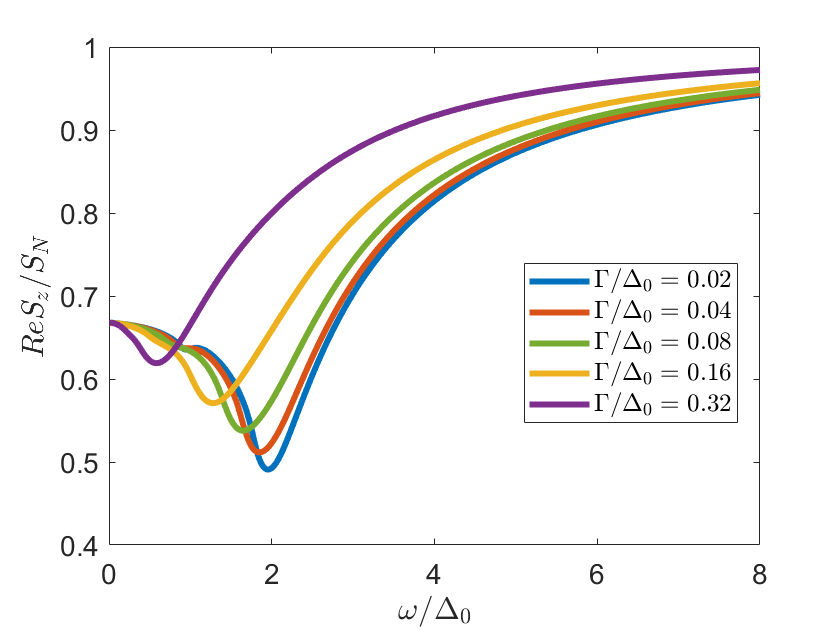}
    \includegraphics[width=8.6cm]{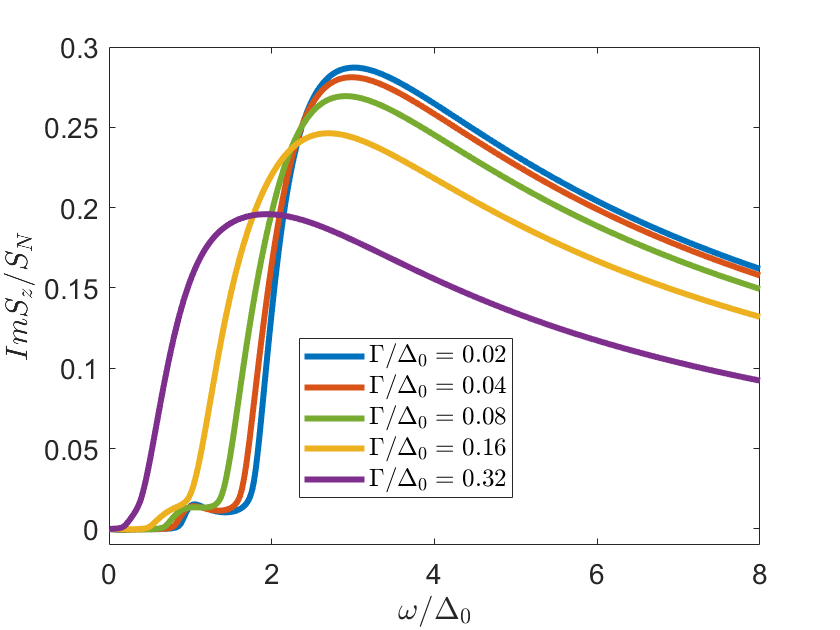}
    \includegraphics[width=8.6cm]{Figures_to_use/A.png}
    \includegraphics[width=8.6cm]{Figures_to_use/B.png}
    \caption{The dependence of the generated spin on the spin relaxation scattering rate at $T = 0.1T_c$ in the limit $d\ll \xi, l_s$. The curves have been plotted with the Dynes parameter $\eta = 0.02\Delta_0$ to ensure numerical stability. The excitation gap depends on $\Gamma$, and is visible from the absence of an out-of-phase component of spin for small frequencies.}
    \label{fig:Gammadep}
\end{figure*}

Next, we consider the influence of magnetic spin relaxation, which has contributions from spin-flip scattering on magnetic impurities and from the Dyakonov-Perel type of relaxation for altermagnets \cite{vasiakin2025disorder}.
If we assume that the magnetic relaxation length $l_s$ is much larger than the thickness of the material, we can still use the same formalism as before, in which the generation of the spin is almost unaffected, but the bulk properties of the material change due to the suppression of superconductivity. The influence of magnetic spin relaxation is similar to that of the Dynes parameter, as illustrated in Fig.~\ref{fig:Gammadep}, especially for the in-phase component, Fig.~\ref{fig:Gammadep}(a). However there are also notable differences. First of all, spin relaxation does not induce a strong spin signal for frequencies below $\omega = 2\Delta (\Gamma)$, but it does for $\omega>2\Delta(\Gamma)$. Next to this, it severely decreases the maximum induced spin and the out-of-phase component of the spin, see Fig.~\ref{fig:Gammadep}(b). This is in line with the expectation that magnetic spin relaxation both suppresses superconductivity and hampers the creation of spin.
For larger thicknesses, spin relaxation plays another important role, because the assumption of an almost linear dependence of the spin accumulation on the vertical coordinate cannot be motivated if $d$ becomes of the order of the spin relaxation length. Indeed, it limits the maximum magnitude of spin, which without relaxation increases indefinitely as $d\xrightarrow{}\infty$, and it limits the distance to the boundary over which the spin can be found to the order of the spin relaxation length $l_s = \sqrt{\frac{D}{\Gamma}}$. This effect is frequency dependent, spin-relaxation is weakest at low frequencies and strongest around $\omega = 2\Delta(\Gamma)$ \cite{bergeret2018colloquium}.

\section{Discussion}

We have shown that although superconducting altermagnets do not display the equilibrium spin-splitter effect, they exhibit both out-of-phase and in-phase spin-splitter conductivities in the presence of time-dependent driving, enabled by time dependence and dissipation, respectively.
 The strength of this nonequilibrium spin-splitter effect and the relative phase between the driving field and the maximum magnitude of the induced time dependent spin depends strongly on frequency. We show that, unlike in the case of the nonequilibrium spin-Hall effect, the out-of-phase component of the spin density in the nonequilibrium spin-splitter effect does not diverge in the adiabatic limit. Instead, it vanishes for frequencies below the gap, due to the absence of dissipation in this regime.
 Moreover, in contrast to the spin-Hall effect, the in-phase component remains finite below the gap. Since the condensate does not transport or carry spin in the absence of equal spin pairs, this signal can be fully attributed to the off-resonance reactive response of the quasiparticles. At zero frequency the in-phase component converges to a finite value. The absence of the divergence is directly linked to the absence of any equilibrium spin-splitter effect \cite{kokkeler2025quantum}. 

The magnitude of the spin-splitter effect is sensitive to the parameters of the system. For example, by tuning the temperature below but not far below $T_c$, the adiabatic in-phase spin accumulation can be 3 times as high as its normal state value, and more than 5 times as high as its low-temperature value. The imaginary part remains vanishing in the adiabatic limit, but close to $T_c$ a peak develops at a low frequency.

We have also  investigated the influence of magnetic spin relaxation and the effect of in-gap quasiparticles through the Dynes parameter, and shown that both alter the spin-splitter effect by reducing the sharpness of its features, but a clear frequency dependence remains as long as the relaxation rate is smaller than $\Delta_0$.  In the presence of in-gap quasiparticles a residual in-gap density of states remains and affects the behavior of the generated spin for frequencies below the gap. Specifically, it gives rise to an out-of-phase contribution even below the gap. For magnetic spin relaxation on the other hand, the main effects are the suppression of the pair potential, which makes all features appear at lower frequencies and the suppression of the maximum generated spin, which is attributed to the inability to create spin.

The predicted effects can be studied either in materials such as \text{RuO}$_2$, where altermagnetism and superconductivity are predicted to coexist \cite{uchida2020superconductivity,ruf2021strain,occhialini2022straininduced,ahn2019antiferromagnetism,smejkal2020crystal,karube2022observation}, or in hybrid superconductor/altermagnet films, where superconducting and magnetic proximity effects enable the coexistence of both phenomena. Measurements can be performed in the same setups
  used to detect the AC spin Hall effect, for example with the help of spin-polarized STM \cite{mukasa1995spin}, using a nonlocal spin valve \cite{hubler2012longrange}, or in a Hall bar setup \cite{sanz2020quantification}. 
With this, our results provide an alternative way to prove the presence of altermagnetism in a material, next to the anomalous Hall effect without stray fields \cite{smejkal2020crystal}, the piezomagnetic effect \cite{aoyama2024piezomagnetic}, direct probes of the band structure such as ARPES measurements \cite{osumi2024observation}, the normal state spin-splitter effect \cite{gonzalez2021efficient} and the proximity induced magnetization \cite{kokkeler2025quantum}.

All in all, our results show that the nonequilibrium spin-splitter effect in altermagnets is strongly frequency dependent and can be tuned by tuning parameters of the system such as temperature. We find that the frequency dependence is distinctively different from that for spin-Hall responses, making it suitable for the distinction between the two effects and an indicator for the altermagnet spin-splitter effect.
\section{Acknowledgements}
The work of T.K. and  T.T. H. was supported by the Research Council of Finland through DYNCOR, Project Number 354735 and through the Finnish Quantum Flagship, Project Number 359240. F.S.B  acknowledges financial support from the Spanish MCIN/AEI/10.13039/501100011033 
through grants 
PID2023-148225NB-C31 and TED2021-130292B-C42, and from the European Union’s Horizon Europe 
through grant JOSEPHINE (No. 101130224).  

\bibliography{sources.bib}
\newpage
\appendix
\onecolumngrid\
\section{Real observables}\label{sec:RealObservables}
In this section we prove the claim made in the introduction that the currents and the spin accumulations in the model we consider in this manuscript are real. To this end we show that the observables $X(\omega)$ satisfies the relation $X(-\omega) = X(\omega)^*$. This motivates our choice to only show positive frequencies in the main text.
To this end, we make use of the charge conjugation symmetry
\begin{align}
    g(t_1,t_2) & = \rho_1\tau_1\sigma_2 g^T(t_2,t_1)\sigma_2\tau_1\rho_1\;,
\end{align}
and the chronology symmetry, which at the saddle point implies
\begin{align}
    g(t_1,t_2) & = -\rho_2 \tau_3 g^\dagger(t_2,t_1)\tau_3\rho_2\;.
\end{align}

Their product gives the following relation on the Green's function:
\begin{align}
    g(t_1,t_2) = -\rho_3\tau_2\sigma_2 g^{*}(t_1,t_2)\tau_2\sigma_2\rho_3\;.
\end{align}

This relation has consequences for the matrix current as well.
For example for the matrix current to first order in the vector potential, we have for normal metals
\begin{align}
    \mathcal{J}_k(\omega,t_1,t_3) &= -D\frac{E_0}{\omega}g(t_1,t_2) \circ (\tau_3 e^{-i\omega t_2}g(t_2,t_3)-g(t_2,t_3)\tau_3 e^{-i\omega t_3})\nonumber\\&
    =-D\frac{E_0}{\omega}\Big((-\rho_3\tau_2\sigma_2g^*(t_1,t_2)\rho_3\tau_2\sigma_2)\circ(\tau_3 e^{-i\omega t_2}(-\rho_3\tau_2\sigma_2g^*(t_2,t_3)\rho_3\tau_2\sigma_2)-(-\rho_3\tau_2\sigma_2g^*(t_2,t_3)\rho_3\tau_2\sigma_2)\tau_3 e^{-i\omega t_3})\Big)\nonumber\\&=
    \rho_3\tau_2\sigma_2 D\frac{E_0}{\omega} g^*(t_1,t_2)\circ(\tau_3 e^{-i\omega t_2}g^*(t_2,t_3)-g^*(t_2,t_3)\tau_3 e^{-i\omega t_3})\rho_3\tau_2\sigma_2\nonumber\\&=-\rho_3\tau_2\sigma_2\Big(-D\frac{E_0}{-\omega}g(t_1,t_2)\circ (\tau_3 e^{i\omega t_2}g(t_2,t_3)-g(t_2,t_3)\tau_3 e^{i\omega t_3})\Big)^*\rho_3\tau_2\sigma_2\nonumber\\& = -\rho_2\tau_2\sigma_2 \mathcal{J}_k(-\omega,t_1,t_3)^{*}\rho_2\tau_2\sigma_2\;.
\end{align}

From this we conclude that the charge current and spin currents satisfy
\begin{align}
    j_k(\omega) &= \text{tr}\tau_3\mathcal{J}_k(\omega,t_1,t_1) = j_k(-\omega)^*\;,\\
    j_k^a(\omega) &= \text{tr}\sigma_a\mathcal{J}_k(\omega,t_1,t_1) = j_k^a(-\omega)^*\;.
\end{align}

For altermagnets we have a contribution from one extra term, which reads, following an analogous derivation
\begin{align}
    \mathcal{J}^{\text{AM}}_k(\omega,t_1,t_3) &= D\frac{E_0}{\omega}PT_{xy}\Bigg(\{\tau_3\sigma_z,g(t_1,t_2)(\tau_3 e^{-i\omega t_2}g(t_2,t_3)-g(t_2,t_3)\tau_3 e^{-i\omega t_3})\}\nonumber\\&+g(t_1,t_2)\circ(\tau_3 e^{-i\omega t_2}g(t_2,t_3)-g(t_2,t_4)\tau_3 e^{-i\omega t_4})\circ g(t_4,t_5)\tau_3\sigma_z \circ g(t_5,t_3)\nonumber\\&+g(t_1,t_4)\tau_3 \sigma_z \circ g(t_4,t_5)\circ g(t_5,t_2)\circ (\tau_3 e^{-i\omega t_2}g(t_2,t_3)-g(t_2,t_3)\tau_3 e^{-i\omega t_3})\Bigg)\nonumber\\&
    =-\rho_3\tau_2\sigma_2\Bigg(D\frac{E_0}{-\omega}PT_{xy}\Big(\{\tau_3\sigma_z,g(t_1,t_2)\circ (\tau_3 e^{-i\omega t_2}g(t_2,t_3)-g(t_2,t_3)\tau_3 e^{-i\omega t_3})\}\nonumber\\&+g(t_1,t_2)\circ(\tau_3 e^{i\omega t_2}g(t_2,t_3)-g(t_2,t_4)\tau_3 e^{i\omega t_4})\circ g(t_4,t_5)\tau_3\sigma_z \circ g(t_5,t_3)\nonumber\\&+g(t_1,t_4)\tau_3 \sigma_z \circ g(t_4,t_5)\circ g(t_5,t_2)\circ(\tau_3 e^{i\omega t_2}g(t_2,t_3)-g(t_2,t_3)\tau_3 e^{i\omega t_3})\Big)\Bigg)^*\rho_3\tau_2\sigma_2\nonumber\\& = -\rho_2\tau_2\sigma_2 \mathcal{J}_k(-\omega,t_1,t_3)^{*}\rho_2\tau_2\sigma_2\;.
\end{align}
From this we see that in altermagnets the currents satisfy exactly the same symmetries and hence also our Usadel equation for altermagnets leads to real currents.

For the spin density we have
\begin{align}
    \partial_y S^z (\omega,t_1,t_1) &= \text{tr}\tau_3\sigma_zg(t_1,t_2)\circ \Big(\mathcal{J}(t_2,t_1)(\omega,t_2,t_1)+\mathcal{J}^{\text{AM}}(\omega,t_2,t_1)\Big) \nonumber\\&= \text{tr}\tau_3\sigma_z\rho_3\tau_2\sigma_2g(t_1,t_2)^*\circ \rho_3\tau_2\sigma_2\Big(\rho_3\tau_2\sigma_2\mathcal{J}(\omega,t_2,t_1)^*\rho_3\tau_2\sigma_2+\rho_3\tau_2\sigma_2\mathcal{J}^{\text{AM}}(-\omega,t_2,t_1)^*\rho_3\tau_2\sigma_2\Big)\nonumber\\& = \text{tr}\tau_3\sigma_zg^*(t_1,t_2)\circ \Big(\mathcal{J}(-\omega,t_2,t_1)^*+\mathcal{J}^{\text{AM}}(-\omega,t_2,t_1)^*\Big) = \partial_y S_z(-\omega,t_1,t_3)^*\;.
\end{align}
This shows that also for spin all physical observables are real.

In the presence of a spin-Hall effect the derivation is similar, but, we only omit the $\tau_3$'s above, add a $g$ and the matrix current reads:
\begin{align}
    \mathcal{J}^{SH}_k(\omega,t_1,t_2) &= -D\frac{E_0}{\omega}\theta\epsilon_{ijz}\sigma_z\Big( e^{-i\omega t_1}g(t_1,t_2)-g(t_1,t_2)  e^{-i\omega t_2}\Big)\nonumber\\&=-D\frac{E_0}{\omega}\theta\epsilon_{ijz}\sigma_z\Big( e^{-i\omega t_1}\rho_3\tau_2\sigma_2g(t_1,t_2)^*\rho_3\tau_2\sigma_2-\rho_3\tau_2\sigma_2g(t_1,t_2)^*\rho_3\tau_2\sigma_2  e^{-i\omega t_2}\Big)\nonumber\\&=\Big(-D\frac{E_0}{-\omega}\theta\epsilon_{ijz}\sigma_z( e^{-i\omega t_1}g(t_1,t_2)-g(t_1,t_2)^*  e^{-i\omega t_2})\Big)^* = J^{SH}_k(-\omega,t_1,t_2)\;.    
\end{align}
which shows that also the spin Hall matrix current, and consequently the spin generated by the spin-Hall effect are real observables. This shows that all observables predicted by our theory are real, and it allows us to focus on the response for positive $\omega$.

\section{Derivation of spin accumulation at the interface}\label{sec:SpinAccumulationInterface}
In this section we derive the expression for the spin accumulation at the interface for a narrow and broad stripe, Eq.~(\ref{eq:Spinacc}).

The spin accumulation at the interface is defined as
\begin{align}
    S^{z} = \text{tr}\tau_3\rho_1\sigma_z g\;.
\end{align}

First we consider a narrow stripe.
Since by symmetry of the problem, the spin is odd in $y$, and hence $S^{z}(d/2)\approx \frac{d}{2}\partial_y S^{z}(d/2)$.
This latter quantity can be expressed using the boundary condition. Indeed, we have
\begin{align}
    0 = J_y = -Dg\hat{\partial}_y g +PT_{xy}\{\tau_3\sigma_z+g\circ\tau_3\sigma_z g\commastsym g\circ\hat{\partial}_x g\}+iPK_{xy}[\tau_3\sigma_z+g\circ\tau_3 \sigma_z g\commastsym\hat{\partial}_jg ]\;.
\end{align}

With this we find

\begin{align}
    S^{z}(d/2) &= PT_{xy}\frac{d}{2}\text{tr}\rho_1\tau_3\sigma_z g\circ\{\tau_3\sigma_z+g\tau_3\sigma_z\circ g\commastsym g\hat{\partial}_xg\} \nonumber\\&= PT_{xy}\frac{d}{2}\text{tr}\Big(\rho_1\tau_3g(E)\tau_3 g(E) (\tau_3g(E+\omega)-g(E)\tau_3)+2\rho_1(\tau_3g(E+\omega)-g(E)\tau_3)\nonumber\\&+\rho_1\tau_3(\tau_3g(E+\omega)-g(E)\tau_3)g(E+\omega)\tau_3g(E+\omega)\Big)\nonumber\\
    &= PT_{xy}\frac{d}{2}\text{tr}\Big(\rho_1\tau_3g(E)\tau_3 g(E) \tau_3g(E+\omega)-\rho_1 g(E)\tau_3+2\rho_1(\tau_3g(E+\omega)-g(E)\tau_3)\nonumber\\&+\rho_1\tau_3 g(E+\omega)-\rho_1\tau_3g(E)\tau_3g(E+\omega)\tau_3g(E+\omega)\Big)\nonumber\\
     &= PT_{xy}\frac{d}{2}\text{tr}\Big(3\rho_1  (\tau_3 g(E+\omega)-g(E)\tau_3)-\rho_1\tau_3 g(E)\tau_3(g(E+\omega)-g(E))\tau_3g(E+\omega)\tau_3\Big)\;.
\end{align}

This is Eq.~(\ref{eq:Spinacc}) in the main text.
\onecolumngrid\

\section{Adiabatic limit}\label{sec:Adiabatic}
In this Appendix we show analytically that the generated spin-accumulation shown in Fig.~\ref{fig:Spin} converges to $\frac{2}{3}S_{z,N}$ when $\omega\xrightarrow{}0$.

In the adiabatic limit ($\omega\ll\Delta$) we may write $g(E+\omega)\approx g(E)+ \omega\partial_E g$. In that case Eq.~(\ref{eq:Spinacc}) to first order in $\omega$ reduces to
\begin{align}
    S_z &= -\frac{\nu_0}{8} dE E_0PT_{xy}E_0\int_{-\infty}^{\infty} \text{tr}3 \rho_1 \tau_3 \partial_E g - \rho_1\tau_3g\tau_3\partial_E g\tau_3 g\nonumber\\&= \frac{S_{z,N}}{32}\int_{-\infty}^{\infty}dE \text{tr}3 \rho_1 \tau_3 \partial_E g - \rho_1\tau_3g\tau_3\partial_E g\tau_3 g\;.
\end{align}

The first integral is easy to evaluate and results in $3 \text{tr}\rho_1(g(E\xrightarrow{}\infty)-g(E\xrightarrow{}-\infty)) = 24$. The second term in the normal state is equivalent to the two terms and results in $8$, but in the superconducting state it is not equivalent and we calculate it below.

Using that for the zeroth order Green's function we have $g^K = (g^R-g^A)f(E)$, we write this term as
\begin{align}
    -\int_{-\infty}^{\infty} dE \text{tr}(\tau_3g^R\tau_3g^R\tau_3g^R-\tau_3g^A\tau_3g^A\tau_3g^A)\partial_E f + (\tau_3g^R\tau_3\partial_E g^R\tau_3g^R-\tau_3g^A\tau_3\partial_E g^A\tau_3g^A)f\;.
\end{align}
At zero temperature, we have $\partial_E f = 2\delta(E)$. Since for a superconductor $\text{tr}\tau_3 g\tau_3 g\tau_3g (E = 0) =  0$, this does not contribute in the superconducting state. The second term can be rewritten as 
\begin{align}
    -\frac{1}{3}\int_{-\infty}^{\infty} dE\text{tr} f(E)\partial_E (\tau_3g^R\tau_3g^R\tau_3g^R-\tau_3g^A\tau_3g^A\tau_3g^A) = -\frac{2}{3}\int_0^{\infty} dE \partial_E (\tau_3g^R\tau_3g^R\tau_3g^R-\tau_3g^A\tau_3g^A\tau_3g^A) = -\frac{8}{3}\;.
\end{align}
Thus, comparing the adiabatic spin in the superconducting state to the normal state, we find
\begin{align}
    \frac{S_z(\omega = 0, \Delta\neq 0)}{S_{z,N}} = \frac{24-\frac{8}{3}}{32}= \frac{2}{3}\;.
\end{align}
This is in agreement with our numerical calculations shown in Fig.~\ref{fig:Spin}.
\end{document}